\documentclass[acmsmall]{acmart}

\usepackage{multirow}
\usepackage{enumitem}
\usepackage{CJK}

\renewcommand{\refname}{REFERENCES}
\renewcommand{\acksname}{ACKNOWLEDGMENTS}
\AtBeginDocument{%
  }

\setcopyright{rightsretained}
\acmJournal{PACMHCI}
\acmYear{2024} \acmVolume{8} \acmNumber{CSCW2} \acmArticle{403} \acmMonth{11}\acmDOI{10.1145/3686942}

\makeatletter
\gdef\@copyrightpermission{
  \begin{minipage}{0.2\columnwidth}
   \href{https://creativecommons.org/licenses/by/4.0/}{\includegraphics[width=0.90\textwidth]{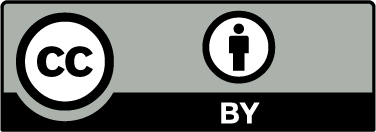}}
  \end{minipage}\hfill
  \begin{minipage}{0.8\columnwidth}
   \href{https://creativecommons.org/licenses/by/4.0/}{This work is licensed under a Creative Commons Attribution International 4.0 License.}
  \end{minipage}
  \vspace{5pt}
}
\makeatother

\begin{document}

\title[Uncovering Hidden Hurdles to Collaborative Writing ]{(Dis)placed Contributions: Uncovering Hidden Hurdles to Collaborative Writing Involving Non-Native Speakers, Native Speakers, and AI-Powered Editing Tools}


\author{Yimin Xiao}
\affiliation{%
  \institution{University of Maryland}
  \city{College Park, MD}
  \country{USA}}
\email{yxiao@umd.edu}

\author{Yuewen Chen}
\affiliation{%
  \institution{University of Maryland}
  \city{College Park, MD}
  \country{USA}}
\email{yc0506@terpmail.umd.edu}

\author{Naomi Yamashita}
\affiliation{%
  \institution{Kyoto University}
  \city{Kyoto}
  \country{Japan}}
\email{naomiy@acm.org}

\author{Yuexi Chen}
\affiliation{%
  \institution{University of Maryland}
  \city{College Park, MD}
  \country{USA}}
\email{ychen151@umd.edu}

\author{Zhicheng Liu}
\affiliation{%
  \institution{University of Maryland}
  \city{College Park, MD}
  \country{USA}}
\email{leozcliu@umd.edu}

\author{Ge Gao}
\affiliation{%
  \institution{University of Maryland}
  \city{College Park, MD}
  \country{USA}}
\email{gegao@umd.edu}

\renewcommand{\shortauthors}{Yimin Xiao et al.}

\begin{abstract}
 Content creation today often takes place via collaborative writing. A longstanding interest of CSCW research lies in understanding and promoting the coordination between co-writers. However, little attention has been paid to individuals who write in their non-native language and to co-writer groups involving them. We present a mixed-method study that fills the above gap. Our participants included 32 co-writer groups, each consisting of one native speaker (NS) of English and one non-native speaker (NNS) with limited proficiency. They performed collaborative writing adopting two different workflows: half of the groups began with NNSs taking the first editing turn and half had NNSs act after NSs. Our data revealed a “late-mover disadvantage” exclusively experienced by NNSs: an NNS’s ideational contributions to the joint document were suppressed when their editing turn was placed after an NS’s turn, as opposed to ahead of it. Surprisingly, editing help provided by AI-powered tools did not exempt NNSs from being disadvantaged. Instead, it triggered NSs’ overestimation of NNSs’ English proficiency and agency displayed in the writing, introducing unintended tensions into the collaboration. These findings shed light on the fair assessment and effective promotion of a co-writer’s contributions in language diverse settings. In particular, they underscore the necessity of disentangling contributions made to the ideational, expressional, and lexical aspects of the joint writing. 
\end{abstract}



\received{July 2023}
\received[revised]{January 2024}
\received[accepted]{March 2024}

\maketitle

\section{INTRODUCTION}
Collaborative writing, the process in which multiple individuals contribute to joint content production by taking sequential turns, is essential in modern educational, academic, and industrial settings [56]. Since at least the early 1990s, CSCW and HCI scholars have been investigating how co-writers coordinate with one another and exploring ways to facilitate this process [3, 50, 63, 80]. Their endeavors have led to the evolution of collaborative writing systems over generations, ranging from annotation-based change trackers used with word processors (e.g., [3, 22, 57]) to visualization systems that capture changes to the document’s content across time (e.g., [83, 86, 87]). However, little of this research was conducted with co-writer groups consisting of both non-native speakers (NNS) and native speakers (NS) of the working language. 

Recent empirical and anecdotal evidence indicates frequent participation of NNSs in today’s collaborative writing practice [15, 71, 92]. Much of this evidence finds language diversity a mixed blessing. On the one hand, studies conducted within English-speaking workplaces and classrooms argue that the involvement of NNSs is likely to benefit content production at the group level [32, 60, 69]. On the other hand, writing in a non-native language can be challenging, especially for individuals with limited proficiency [16, 33]. There remains a pressing yet unresolved question about how to better activate the potential of NNSs as well as co-writer groups involving them. 

The current research aims to understand and enhance collaborative writing between NNSs and NSs. Inspired by the theory of footing [29], we suspect that NNSs’ lack of working language proficiency can limit their expression of ideas; however, it does not necessarily affect their participation in ideation per se. Thus, effective coordination between NNSs and NSs should alleviate the former’s concerns in editing the expressional aspect of the joint document, while encouraging their input into the ideational aspect. Careful planning of the turn-taking order between NNSs and NSs may guide co-writer groups toward this goal. 

We conducted online experiments with 32 NNS-NS dyads to examine the effect of turn-taking order on collaborative writing. Half of the co-writer groups began turn-taking with NNSs, while NSs took the first editing turn in the other half. We found that NNSs were more likely to contribute to the ideational aspect of the joint document when they took the editing turn ahead of NSs, as opposed to after NSs. Additionally, participants perceived better task experiences when the turn-taking began with NNSs. This perception was evident not only in the ratings provided by NSs but also in both co-writers’ interview responses. 

Moreover, we inquired about the role of AI-powered editing tools in collaborative writing between NNSs and NSs. While these tools hold the promise of closing the expression-ideation gap experienced by NNSs, our data revealed several risks undertaken by NNSs when turned to these tools for editing suggestions. We also identified instances where the use of AI-powered tools by NNSs disrupted interpersonal dynamics between co-writers in unexpected ways. 

Findings from our research advance the empirical understanding of collaborative writing in contexts featuring language diversity. They also generate insights for future system design that promotes the inclusion of co-writers with different language backgrounds, as well as the equitable assessment of each party’s contributions. 

\section{RELATED WORK}
In this section,
we begin with literature indicating how NNSs differ from NSs in the production of written content (Section 2.1). We then introduce the notion of footing, which provides a unique lens to examine a co-writer’s contributions to the joint document from two distinct aspects: expression and ideation (Section 2.2). Following this, we explore possible ways to alleviate the constraints faced by NNSs who participate in collaborative writing with NSs, guided by prior work considering NS-NNS interactions under various workflows in terms of the turn-taking orders (Section 2.3), as well as emerging discussions considering writing with AI-powered tools (Section 2.4). 

\subsection{Native vs. Non-Native Speakers in the Production of Lexical Content  }

Content production in a non-native language is not easy. In the case of document writing, decades of research have illustrated the struggles experienced by NNSs. 

One set of studies draws conclusions through comparisons between writing performed in non-native and native languages by the same person. In particular, Chenoweth and Hayes conducted controlled experiments with NSs of English, all of whom were learning French or German as their non-native language [16]. Analysis of each person’s writing process revealed that participants paused twice as much when writing in French or German compared to English. The length of their essays generated in the non-native language was significantly shorter than in English. Wolfersberger performed case studies with NSs of Japanese who wrote in English as a non-native language [89]. Participants reported difficulties in transferring strategies from Japanese writing into English writing and, as a result, produced English essays with limited content. This contrast echoes Uzawa and Cumming’s previous finding that individuals tended to reduce the amount and complexity of content in a document when writing as an NNS [84]. 

Another line of research compares the lexical features of writing outcomes generated by NNSs and NSs as two independent groups. For instance, Ferris collected English essays written by NSs and NNSs of English, all following the same prompts [20]. NNSs produced significantly shorter essays with less variety of language use than NSs’ essays. Severino et al. analyzed a large volume of requests submitted to an online English writing center by individuals with various language backgrounds [75]. Their data showed that, NNSs were much more concerned about the lexical content of their writing than NSs of English. They frequently sought help for the improvement of word choices and syntactic structures in English. 

In short, the above literature suggests that lack of proficiency in a language can limit the production of written content in that language. While the bulk of empirical evidence has been collected in the context of individual writing, we hypothesize that a similar relationship between language proficiency and content production also applies to collaborative writing: 

\begin{itemize}
    \item[] \textbf{\textit{H1.}} NNSs will make fewer edits to the lexical content of a joint document than NSs.
\end{itemize}

\subsection{Non-Native Speaker’s Potential from the Lens of Footing  }

Emphasizing a co-writer’s ability to “produce more lexical edits” forebodes a pessimistic future for NNSs as well as the NSs working with them. Since language proficiency cannot be boosted significantly within a short period of time, there is only a slim possibility to promote an NNS’s volume of lexical edits to an equal level as an NS’s. In this section, we introduce the notion of footing as a different lens to reconsider the contributions that NNSs and NSs can bring to their collaborative writing. Through this lens, a person’s contribution is examined against the function of their content production rather than their volume of lexical edits. 

According to Goffman, footing describes a person’s relation to the lexical content they produce. This relationship can happen in more than one format [29]. In particular, a person can lead the decision about how to word the expression of a message; they can also contribute to the underlying idea of the message. The expressional and ideational aspects of a given piece of lexical content are often managed by the same person but not always. 

NNS-NS interactions constitute a scenario where the people responsible for the expressional and ideational aspects of a message can be separated. For example, Hosoda analyzed the corpus of verbal communication between NNSs and NSs of Japanese [34]. Their data revealed that it was common for NSs to rework the expression of NNSs’ ideas to ensure the clarity of the message and prevent misunderstanding. Kurhila examined the corpus of NS-NNS communication in Finnish [51]. They found that NNSs often experienced uncertainty when attempting to articulate their own ideas. When NSs sensed this uncertainty, they offered corrections to improve NNSs’ initial expressions. In global work using English as a common language, prior research has documented extensive cases where NNSs of English chose to withdraw themselves from idea exchange at big meetings due to anxiety about linguistic expression [69, 81]. Their NS colleagues often acted as “the advocate,” presenting the collective thoughts of the whole group in English [59, 62]. 

The main takeaway from the above literature is that a person’s contribution to content production can be separated into two distinct aspects: expression and ideation. These two aspects appear to impose different constraints on NNSs. In the case of collaborative writing, NNSs’ ability to edit the expressional aspect of written content is directly hindered by their limited working language proficiency. NNSs’ ability to originate or elaborate ideas for the benefit of the co-writer group is not affected by their language proficiency; however, they may feel demotivated to perform ideation due to anticipated expression challenges. Thus, to fulfill NNSs’ potential, task coordination should be planned so as to alleviate NNSs’ concerns about editing the expressional aspect of a joint document, while encouraging their input into its ideation. 

\subsection{Expressional Edits, Ideational Edits, and the Workflow Connecting Co-Writers }
The workflow, in terms of how co-writers arrange the order of their editing turns, constitutes a crucial component of task coordination at the group level. Our literature review suggests that the likelihood of an NNS editing the expressional or ideational aspects of a joint document may both vary according to their order of taking the editing turn against an NS’s. 

Specifically, previous studies have found that it is challenging for individuals to edit expressions produced in their non-native language. As demonstrated in Section 2.2, NNSs in oral conversations often request help from NSs to improve messages initiated by themselves [34, 51]; however, reports of NNSs reworking NSs’ oral messages are scant. In writing, it is a common observation that, although NNSs can be highly aware of the possible issues with their written expressions, they lack the skill to perform edits independently [21, 27]. 

A more recent line of research has examined whether corrective feedback on an early version of NNSs’ writing promotes their edits at subsequent times. For instance, Karim and Nassaji compared the editing behavior of NNSs receiving no feedback with those who received detailed explanations of their expression issues in early drafts [42]. They found that direct feedback at the early stage increased NNSs’ chances of making appropriate revisions to their language use at a later stage. Kang examined language modeling as an alternative way to elicit NNSs’ edits [41]. Participants in their study were divided into two groups: half received no feedback on their writing, while the other half received model texts written by NSs on the same topic. Analysis of subsequent writing suggested that model texts primed NNSs to reflect on differences between the models and their own writing and, subsequently, make revisions to their initial expressions. 

Based on this literature, we hypothesize that exposure to an NS’s edits made at an earlier turn will increase the likelihood of an NNS editing the expressional aspect of the document; in contrast, being assigned to an earlier position in the turn-taking will provide NNSs fewer opportunities to perform such edits, as there are fewer clues to follow: 

\begin{itemize}
    \item[] \textbf{\textit{H2.}} NNSs will be more likely to edit the expressional aspect of a joint document when they take editing turns after NSs, as opposed to ahead of NSs. 
\end{itemize}

Meanwhile, when the wording of an earlier draft has been extensively edited by others, people may avoid elaborating on the ideas discussed in their writing. This phenomenon has been documented by much previous research (e.g., [45, 50, 77, 78, 94]). As one example, Kepner analyzed the essays written by American college students using Spanish as their non-native language [45]. Some students received teacher’s edits that focused on word choices and grammar at a sentence level, whereas others only received high-level comments on their writing as a whole. The ideational quality of the latter group’s final essays outperformed that of the former’s. Sheppard reported a similar study with NNSs of English [77]. They found that early corrections to NNSs’ language use often discouraged NNSs from adding complexity to their thoughts during revisions. One explanation for these findings is that taking a more conservative attitude toward idea expansion helps NNSs control the anticipated effort for additional language editing by themselves or others [82]. 

Transitioning from individual writing to collaborative work, we hypothesize that exposure to an NS’s edits made at an earlier turn will decrease the likelihood of an NNS editing the ideational aspect of a joint document; in contrast, being assigned to an earlier position in the turn-taking will enable NNSs to concentrate more on ideation:

\begin{itemize}
    \item[] \textbf{\textit{H3.}} NNSs will be more likely to edit the ideational aspect of a joint document when they take editing turns ahead of NSs, as opposed to after NSs. 
\end{itemize}

Notably, previous research, as we have reviewed above, implies that each particular workflow or turn-taking order can better activate NNSs’ potential for one aspect of their content production but limit the other. Given that an ideal setup of the task coordination should encourage co-writers’ contributions to both the expression and ideational aspects of joint writing, we wonder: 

\begin{itemize}
    \item[]\textbf{\textit{RQ1.}} How will co-writers perceive the quality of their task coordination at the end of the task process? In particular, will this perceived quality vary according to the language background of each co-writer and/or the order of turn-taking between co-writers?
\end{itemize}

\begin{itemize}
    \item[] \textbf{\textit{RQ2.}} How will co-writers perceive the value of being able to speak English or a different native language (i.e., Japanese) in the current task context? In particular, will this perception vary according to the language background of each co-writer and/or the order of turn-taking between co-writers?
\end{itemize}

\subsection{Expression and the Use of AI-Powered Editing Tools  }
A side question asked in the current research considers co-writers’ use of AI-powered editing tools for assistance. A growing body of literature in HCI and CSCW has presented cases where people leverage those tools to address the grammar errors in their writing [48, 67], paraphrase sentences [7, 93], translate text across languages [26, 35], and even generate new text in response to human-written prompts [23, 43]. For NNSs, the continuous advancement of AI-powered tools outlines a promising future in which they hold an equal footing to NSs in content production. 

Nevertheless, a small set of recent work has reported preliminary evidence challenging this premise. In an interview study conducted by Kim et al., NNSs of English frequently expressed difficulties in assessing the quality of paraphrased text generated by AI-powered tools, such as Quillbot [49]. Another study by Ito and colleagues asked NNSs of English to write essays in a lab setting and recorded the entire task process [38]. Analysis of the video data indicated that NNSs often turned to AI-powered paraphrasers and translation tools (e.g., Google Translate) when struggling to generate sentences in English. However, they spent significant time scrutinizing the outputs from those tools and were not confident in their decisions to adopt or reject the suggested texts. This literature prompted us to inquire about the possible role of AI-powered tools in collaborative writing between NNSs and NSs: 

\begin{itemize}
    \item[] \textbf{\textit{RQ3.}} How will co-writers make use of AI-powered tools during the task process, if at all? In particular, will NSs and NNSs interpret the value of those tools differently? 
\end{itemize}

\section{METHOD}
We conducted an online experiment with 32 co-writer groups. Each group consisted of one NS of English and one NNS who spoke Japanese as their native language. All groups performed their task using Google Docs. The task required NNSs and NSs to produce content for their joint English document by taking successive editing turns. We manipulated the arrangement of editing turns between co-writers, which resulted in two orders of turn-taking at the group level. We chose this task design after careful considerations of its impact on the internal and external validity of our research findings. Section 7.1 detailed these reflections. 


\subsection{Participants}

Sixty-four individuals participated in the current research. Half were recruited from a university in the United States (N = 32; 18 females, 14 males), all of whom were NSs of English; their mean age was 24.91 (S.D. = 4.38). The remaining half were recruited from a university in Japan (N = 32; 24 females, 8 males), all of whom spoke Japanese as their only native language. Their mean age was 23.84 (S.D. = 3.15). Their self-identified English level was “limited working proficiency,” according to the ILR scale [37]. 

\subsection{Task Procedure}

\subsubsection{Task context and writing prompts.} Our task required all co-writer groups to act as if they were guest writers for \textit{Tech Society}, a pseudo international magazine that provides the public with information about the role of technology in modern life. The magazine features an advice column that publishes articles in response to questions asked by anonymous readers. Co-writer groups were tasked with drafting an English article to address the readers’ questions. A similar setup has been widely used in previous research on collaborative writing [5, 9, 24]. 

We prepared an initial pool of writing prompts containing five questions, each indicating a different writing topic. Prior to the formal task, participants were instructed to read all five questions and identified the one(s) for which they possessed sufficient knowledge to provide a response. We then matched individual participants to form co-writer groups consisting of one NS and one NNS, based on writing topics that both co-writers were knowledgable enough to write about. Three topics remained in our task material following the above process (Appendix A).

Participants were required to communicate their opinions to the readers and support those opinions with concrete evidence. We assigned a word limit of 500-1000 words to each group’s final article, guiding participants to prioritize the quality of their writing rather than its length. 

\subsubsection{\textit{Task procedure and the arrangement of editing turns.} }We provided all participants with preset Google accounts and links to blank Google Docs files for their collaborative writing. NNSs and NSs of the same co-writer group had no personal acquaintance prior to the task, but they had been informed about the language background of the co-writer. 

Every participant began the formal task with a planning turn (i.e., turn\textbf{\textsubscript{0}}), where they individually drafted content to bring into the joint document without consulting their co-writer. This step ensured that all co-writers had actively thought about the task, minimizing the possibility of them acting as free riders in the collaboration.

At the first turn of collaborative writing (turn\textbf{\textsubscript{1}}), one of the co-writers reviewed all the planning content and generated the initial version of the group’s joint document. The document was then passed to the other co-writer for editing (turn\textbf{\textsubscript{2}}). This exchange of editing turns was repeated two more times (turn\textbf{\textsubscript{3 }}and turn\textbf{\textsubscript{4}}) before the group submitted their final article to the researchers. 

To accommodate the time difference across locations as well as the daily schedule of each person, each editing turn took place at the corresponding person’s self-selected hour of their day and place. Each turn or writing session lasted for up to 60 minutes. Half of the co-writer groups performed the task with NNSs taking turn\textbf{\textsubscript{1}}, while NSs took turn\textbf{\textsubscript{1}} in the other half (Figure 1). 

\begin{figure}[h]
  \centering
  \includegraphics[width=\linewidth]{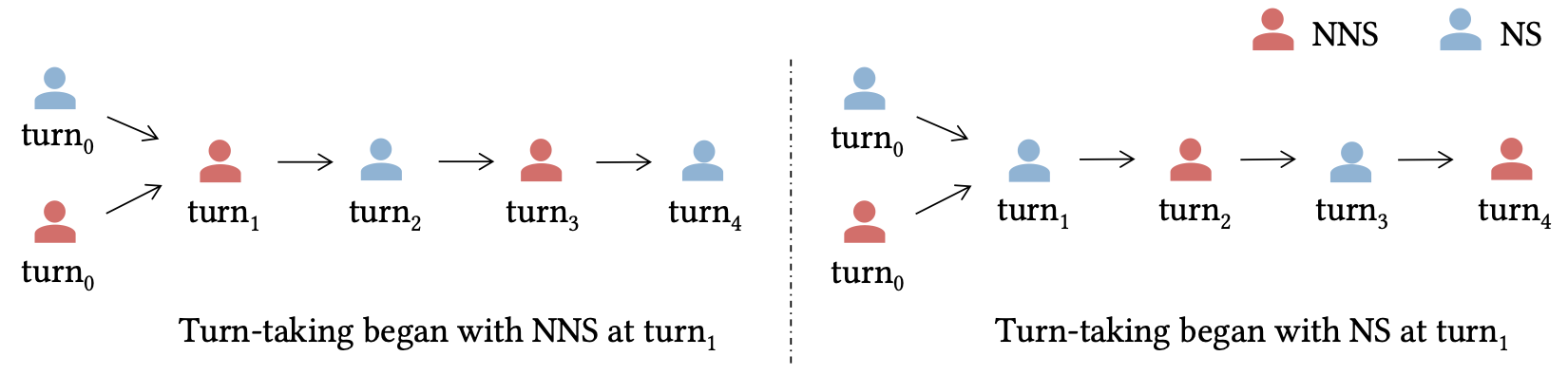}
  \caption{Order of Turn-Taking with NNSs or NSs at Turn\textsubscript{1 }}
\end{figure}

All participants were required to self-record their writing sessions to assist their recall at post-task interviews. These recordings also helped the research team confirm that participants had been adhering to the task instructions. In addition, participants filled out a short Qualtrics survey immediately after the last editing turn of their collaborative work. This survey collected each co-writer’s numerical ratings of their task experience across various aspects.

\subsection{Measurements}

\subsubsection{Changes in the document’s lexical content between every two adjacent turns.} We adopted the Jaccard similarity coefficient to measure the lexical relationship between the documents produced by each group after two adjacent turns. This coefficient was calculated using the following formula: J (A, B) = |A$\cap$B| / |A$\cup$B| , where A and B each represents the unique uni-grams in one of the documents being compared [64]. The value ranges from 0 to 1; a value of 0 indicates no similarity between the documents, whereas a value of 1 indicates the two documents are identical. For the purpose of our research, we used the Jaccard coefficient’s complementary value to measure the lexical distance between the documents: Lexical distance = 1 - J (A, B). A higher value indicates a greater extent of changes in the document’s lexical content. 

\subsubsection{Edits to the expressional and/or ideational aspects of the written content.} We conducted manual coding to identify the expressional and/or ideational edits occurring at each editing turn (see Appendix B for examples), categorized as either “no” or “yes.” Expressional edits referred to changes made to rectify grammar errors, update word choices, or adjust the flow of a given piece of content; ideational edits considered changes made to elaborate, refine, or redirect the meaning conveyed through a given piece of content. Coding was performed by two people blind to our hypotheses, with an initial intercoder reliability of .88. Their task was proceeded as below:

\begin{itemize}
    \item Collecting all joint documents: We gathered documents produced by all the 32 co-writer groups. Each group produced 4 versions of their document across four editing turns, which resulted in at a total number of 128 documents to be coded.
\end{itemize}

\begin{itemize}
    \item Extracting rhetorical pieces within each document: For each document in the above pool, the coders extracted all rhetorical pieces that contained evidence to support the co-writers’ opinions. Each extracted rhetorical piece contained one unique piece of evidence. Notably, this step did not consider the meta content of a document, such as opening and closing sentences with which the co-writers’ greet, show empathy with, or express acknowledgement to the readers. Previous research has suggested that meta content is less relevant to the communication of concrete thoughts [36].
\end{itemize}

\begin{itemize}
    \item Identifying the edits made to each rhetorical piece: For each rhetorical piece, the coders compared its content by the end of an editing turn against that from the previous turn.
\end{itemize}

\subsubsection{Perceived quality of task coordination.} We used 7-point scales to measure each participant’s perceived quality of task coordination between co-writers. This measurement was adopted from established scales invented by Liu and colleagues [55] (e.g., “The coordination between me and my co-writer was effective in general”, Cronbach’s $\alpha$= .78). A higher average rating across the scales indicated a better quality of task coordination.

\subsubsection{Perceived value of being able to speak English and/or Japanese.} We used 7-point scales to measure each participant’s perceived value of being able to speak English and/or Japanese. This measurement was developed based on Neeley and colleagues’ research [61]. Given the difference in language backgrounds between NNSs and NSs, we tailored the wording of the question depending on the type of participants:

\begin{itemize}
    \item Perceived value of being able to speak English: “Speaking English as my non-native language diminished my ability to perform well in this task” [for NNSs]; “Speaking English as my native language enabled me to perform well in this task” [for NSs]. A higher rating on this scale indicated a greater value assigned to the person’s English ability;
\end{itemize}

\begin{itemize}
    \item Perceived value of being able to speak Japanese: “Being able to speak Japanese in addition to English benefited me during this task” [for NNSs]; “Not being able to speak Japanese disadvantaged me during this task” [for NSs]. A higher rating on this scale indicated a greater value assigned to the person’s Japanese ability.
\end{itemize}

\subsubsection{Collaborative writing experience over the entire task process.} We conducted a semi-structured interview with each participant following the completion of the collaborative writing task. All interview sessions took place over Zoom and lasted for about 40 minutes. During the interview, participants were prompted to reflect on a) their individual contributions to the collaborative work, b) the experience of coordinating with their co-writers, and c) the experience of using AI-powered editing tools during the task process, if applicable. We encouraged participants to review recordings of their writing sessions both before and during the interview sessions, which helped their recall of details. All interviews were conducted in the participant’s native language. 

\section{QUANTITATIVE RESULTS}
We performed statistical tests to verify the three hypotheses (H1, H2, H3) and answer two of the open questions (RQ1, RQ2). Preliminary analyses indicated that our dependent variables of interest did not vary according to co-writer’s demographic attributes (i.e., age and gender) other than their language background. Thus, we do not discuss these demographic variables in the formal results. The writing topic did show effects on some of the dependent variables. We considered this factor in all the parametric models by converting the source variable of the writing topic into two binary dummy variables (i.e., whether the task topic was remote learning, whether the task topic was digital privacy) and setting them as control variables. 

\subsection{Edits of the Lexical Content of the Document as a Whole }

\subsubsection{Changes in the document’s lexical content from turn\textbf{\textsubscript{1 }}to turn\textbf{\textsubscript{4}}.} Our H1 predicted that NNSs would make fewer edits to the lexical content of the joint document than NSs. This hypothesis was fully supported by tracking changes in the document’s lexical content after each editing turn since the previous turn (Figure 2).  

\begin{figure}[h]
    \centering
    \includegraphics[width=0.8\linewidth]{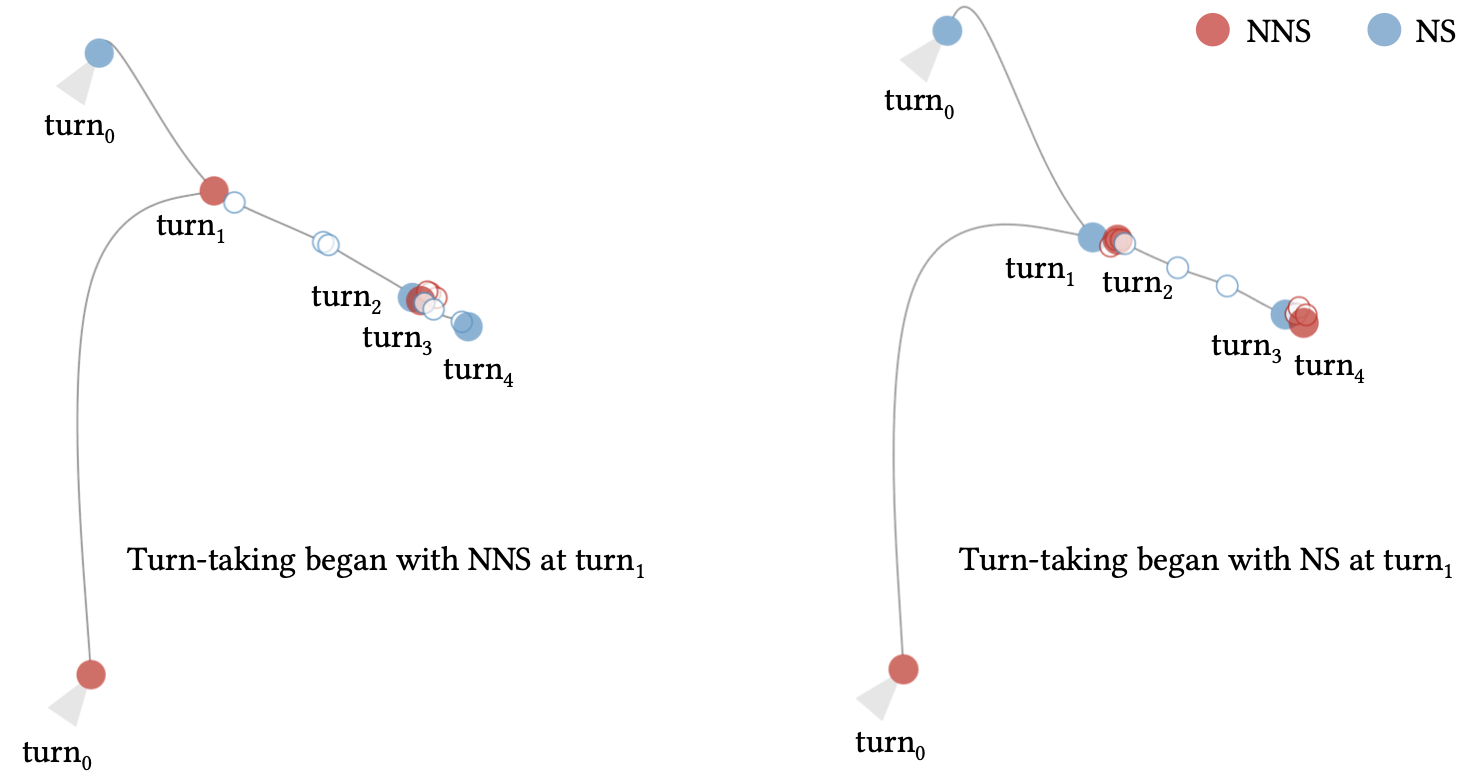}
    \caption{\centering Time Curves [2] that indicate changes in the document’s lexical content across turns. On each curve, the spatial proximity between two dots represents the lexical distance between two corresponding document versions. All versions of the same group’s document are connected by one curve, ordered by the sequence of turn-taking. Each solid dot represents one document version by the end of a given co-writer’s editing turn. Each hollow dot with an empty interior represents an intermediate document version generated during the editing turn. The spatial distribution of dots appears highly similar among groups following the same order of turn-taking. Thus, we present the curves for one randomly selected group whose turn-taking began with an NNS at turn\textsubscript{1 }and another whose turn-taking began with an NS at turn\textsubscript{1}.}
\end{figure}

Specifically, we conducted a 2 $\times$ 3 Mixed Model ANOVA to test H1. Our dependent variable was the value of lexical distance between the same co-writer group’s documents at two adjacent turns. The first independent variable in this model considered the order of turn-taking (order: turn\textbf{\textsubscript{1 }}taken by an NNS or an NS). The second independent variable considered at which specific turn the lexical distance was calculated against its previous turn (turn: turn\textbf{\textsubscript{2, }}turn\textbf{\textsubscript{3, }}or turn\textbf{\textsubscript{4}}). The Huynh-Feldt correction was applied to this test, as the assumption of sphericity was not met. 

The results revealed no significant main effect of the turn: F [1.85, 51.69] = .26, \textit{p} = .75. There was a significant main effect of the order: F [1, 28] = 6.05, \textit{p} < .05. The interaction effect between the turn and the order also appeared to be significant: F [1.85, 51.69] = .25.98, \textit{p} < .01. 

We then performed pairwise comparisons with the Bonferroni correction to examine the detailed change of lexical distance under each order of turn-taking. When the turn-taking started with an NNS at turn\textbf{\textsubscript{1}}, the lexical distance resulting from NNSs’ editing at turn\textbf{\textsubscript{3 }}(M = .04, S.E. = .01) was significantly smaller than the distance resulting from NSs’ editing at turn\textbf{\textsubscript{2 }}(M = .22, S.E. = .03; \textit{p} < .01) and NSs’ editing at turn\textbf{\textsubscript{4 }}(M = .10, S.E. = .02; \textit{p} < .05). When the turn-taking started with an NS at turn\textbf{\textsubscript{1}, }the lexical distance resulting from NSs’ editing at turn\textbf{\textsubscript{3 }}(M = .12, S.E. = .01) was significantly larger than the distance resulting from NNSs’ editing at turn\textbf{\textsubscript{2 }}(M = .04, S.E. = .03; \textit{p} < .05) and NNSs’ editing at turn\textbf{\textsubscript{4 }}(M = .05, S.E. = .02; \textit{p} < .01). 

Together, the above results indicated that NNSs’ edits introduced less change to the lexical content of the joint document compared to NSs’ edits. This contrast held true regardless of a group’s turn-taking order and at which specific turn the lexical distance was calculated. 

\subsubsection{Comparisons between the document at turn\textbf{\textsubscript{1 }}and each co-writer’s planning text at turn\textbf{\textsubscript{0}}.} Besides the above test of H1, we performed additional analysis to explore the relationship between each co-writer group’s document at turn\textbf{\textsubscript{1 }}and each individual’s planning text at turn\textbf{\textsubscript{0 }}(Figure 2).

We conducted a 2 $\times$ 2 Mixed Model ANOVA, setting the value of lexical distance between the document at turn\textbf{\textsubscript{1}} and the planning text at turn\textbf{\textsubscript{0}} as the dependent variable. The first independent variable in this model considered the order of turn-taking (order: turn\textbf{\textsubscript{1}} taken by NNS or NS). The second independent variable considered the author of the planning text (planning text: turn\textbf{\textsubscript{0}}’s text written by an NNS or an NS). 

The results revealed a significant main effect of which group member’s planning text was used for the distance calculation: (F [1, 28] = 26.04, \textit{p} < .01). The main effect of the order appeared to be significant (F [1, 28] = 13.04, \textit{p} < .01), but the interaction effect between these two independent variables was not significant (F [1, 28] = .98, \textit{p} = .33). 

Pairwise comparisons with the Bonferroni correction showed that the joint document at turn\textbf{\textsubscript{1 }}was always more distant from NNSs’ planning text at at turn\textbf{\textsubscript{0 }}(M = .72, S.E. = .02) than from NSs’ (M = .32, S.E. = .03; \textit{p} < .01). This contrast held true regardless of which order of turn-taking was followed by the co-writer group. Meanwhile, the joint document at turn\textbf{\textsubscript{1 }}was always more distant from both of its planning texts at turn\textbf{\textsubscript{0 }}when turn\textbf{\textsubscript{1 }}was taken by an NS (M = .55, S.E. = .01) instead of an NNS (M = .48, S.E. = .01; \textit{p} < .01). 

\subsection{Edits to the Expressional Aspect of the Document  }
\subsubsection{Comparison between the likelihoods of NNSs making edits across different positions.} Our H2 predicted that an NNS would be more likely to edit the expressional aspect of the joint document when they took an editing turn after an NS, rather than ahead of an NS. We tested this hypothesis by examining the likelihoods of the target edits occurring across different positions taken by NNSs. Each position referred to one combination between the order of turn-taking (order: turn\textbf{\textsubscript{1 }}taken by an NNS or an NS) and the turn (turn: turn\textbf{\textsubscript{1, }}turn\textbf{\textsubscript{2, }}turn\textbf{\textsubscript{3, }}or turn\textbf{\textsubscript{4}}). The results rejected H2. 

Specifically, we performed a chi-square test to evaluate the relationship between an NNS’s position and edits occurring in the expressional aspect of each rhetorical piece. The variable of position contained four levels: turn\textbf{\textsubscript{1 }}ahead of its next turn\textbf{\textsubscript{2 }}taken by an NS, turn\textbf{\textsubscript{2 }}after its prior turn\textbf{\textsubscript{1 }}taken by an NS, turn\textbf{\textsubscript{3 }}ahead of its next turn\textbf{\textsubscript{4 }}taken by an NS, turn\textbf{\textsubscript{4 }}after its prior turn\textbf{\textsubscript{3 }}taken by an NS. The variable of edits contained two levels: no vs. yes. 

The results indicated no significant relationship between the two variables: $\chi$\textsuperscript{2} (3, \textit{N} = 296) = 4.94, \textit{p} = .18. As demonstrated in Table 1, the likelihood of an NNS making expressional edits to each rhetorical piece appeared low across all positions. 

\begin{small}
\begin{table}[h]
    \centering
\caption{NNSs’ Edits to the Expressional Aspect of the Rhetorical Pieces }
    \begin{tabular}{p{6cm} p{2.5cm} p{2.5cm} p{0.8cm} }
\toprule
    \multirow{2}{6cm}{NNSs’ position in terms of the combination between the order and the turn} 
        & \multicolumn{3}{l}{Rhetorical pieces appeared in the joint document} \\
        \cline{2-4} & Not edited by NNSs & Edited by NNSs & Total \\
           \midrule

         Turn\textbf{\textsubscript{1}} ahead of its next turn\textbf{\textsubscript{2}} taken by an NS&  52 (expected: 54.81)&  7 (expected: 4.91)& 59 \\  
         Turn\textbf{\textsubscript{2}}  after its prior turn\textbf{\textsubscript{1}} taken by an NS&  70 (expected: 66.89)&  2 (expected: 5.11)& 72 \\  
         Turn\textbf{\textsubscript{3}} ahead of its next turn\textbf{\textsubscript{4}} taken by an NS&  72 (expected: 70.61)&  4 (expected: 5.39)& 76 \\ 
         Turn\textbf{\textsubscript{4}}  after its prior turn\textbf{\textsubscript{3}} taken by an NS&  81 (expected: 82.69)&  8 (expected: 6.31)& 89 \\
        \midrule
         Total&  275&  21& 296 \\
            \bottomrule
\end{tabular}
\end{table}
\end{small}

\subsubsection{Comparison between the likelihoods of NSs making edits across different positions.} In addition to the test of H2, we performed another chi-square analysis to evaluate the relationship between an NS’s position and edits occurring in the expressional aspect of each rhetorical piece. The variable of position contained four levels: turn\textbf{\textsubscript{1}} ahead of its next turn\textbf{\textsubscript{2}} taken by an NNS, turn\textbf{\textsubscript{2}}  after its prior turn\textbf{\textsubscript{1}} taken by an NNS, turn\textbf{\textsubscript{3}} ahead of its next turn\textbf{\textsubscript{4}} taken by an NNS, turn\textbf{\textsubscript{4}} after its prior turn\textbf{\textsubscript{3}} taken by an NNS. The variable of edits contained two levels: no vs. yes. 

The results indicated a significant relationship between the two variables: $\chi$2 (3, N = 291) = 10.66, p < .05. As demonstrated in Table 2, the likelihood of an NS making expressional edits to each rhetorical piece appeared high across positions. The observed counts were higher than the expected counts at the first two positions but lower than them at the last two positions.

\begin{small}
\begin{table}[h]
    \centering
\caption{NSs’ Edits to the Expressional Aspect of the Rhetorical Pieces }
    \begin{tabular}{p{6cm} p{2.5cm} p{2.5cm} p{0.8cm} }
\toprule
    \multirow{2}{6cm}{NSs’ position in terms of the combination between the order and the turn}& \multicolumn{3}{l}{Rhetorical pieces appeared in the joint document} \\
        \cline{2-4} & Not edited by NSs& Edited by NSs& Total\\
        \midrule
         Turn\textbf{\textsubscript{1}} ahead of its next turn\textbf{\textsubscript{2}} taken by an NNS&  25 (expected: 32.81)&  37 (expected: 29.19)& 62
\\  
         Turn\textbf{\textsubscript{2}}  after its prior turn\textbf{\textsubscript{1}} taken by an NNS&  28 (expected: 33.87)&  36 (expected: 30.13)& 64
\\  
         Turn\textbf{\textsubscript{3}} ahead of its next turn\textbf{\textsubscript{4}} taken by an NNS&  47 (expected: 40.75)&  30 (expected: 36.25)& 77
\\ 
         Turn\textbf{\textsubscript{4}}  after its prior turn\textbf{\textsubscript{3}} taken by an NNS&  54 (expected: 46.57)&  34 (expected: 41.43)& 88
\\
        \midrule
         Total&  154&  137& 291
\\
    \bottomrule
\end{tabular}
\end{table}
\end{small}

\subsection{Edits to the Ideational Aspect of the Document }
\subsubsection{Comparison between the likelihoods of an NNS making edits across different positions.} Our H3 predicted that an NNS would be more likely to edit the ideational aspect of the joint document when they took an editing turn ahead of an NS, as opposed to after an NS. We tested it by examining the likelihoods of the target edits occurring across different positions taken by NNSs. Each position referred to one combination between the order of turn-taking (order: turn\textbf{\textsubscript{1 }}taken by an NNS or an NS) and the turn (turn: turn\textbf{\textsubscript{1, }}turn\textbf{\textsubscript{2, }}turn\textbf{\textsubscript{3, }}or turn\textbf{\textsubscript{4}}). The results partially supported H3. 

Specifically, we performed a chi-square test to evaluate the relationship between an NNS’s position and edits to in the ideational aspect of each rhetorical piece. The variable of position contained four levels: turn\textbf{\textsubscript{1 }}ahead of its next turn\textbf{\textsubscript{2 }}taken by an NS, turn\textbf{\textsubscript{2 }}after its prior turn\textbf{\textsubscript{1 }}taken by an NS, turn\textbf{\textsubscript{3 }}ahead of its next turn\textbf{\textsubscript{4 }}taken by an NS, turn\textbf{\textsubscript{4 }}after its prior turn\textbf{\textsubscript{3 }}taken by an NS. The variable of edits contained two levels: no vs. yes. 

The results indicated a significant relationship between the two variables: $\chi$\textsuperscript{2} (3, \textit{N} = 296) = 16.95, \textit{p} < .01. The likelihood of an NNS making ideational edits to each rhetorical piece varied across positions (Table 3). The observed counts for edits were higher than the expected counts at the first position where an NNS took turn\textbf{\textsubscript{1 }}ahead of an NS’s turn\textbf{\textsubscript{2, }}but not at the other positions.  

\begin{small}
\begin{table}[h]
    \centering
\caption{NNSs’ Edits to the Ideational Aspect of the Rhetorical Pieces }
    \begin{tabular}{p{6cm} p{2.5cm} p{2.5cm} p{0.8cm} }
\toprule
    \multirow{2}{6cm}{NNSs’ position in terms of the combination between the order and the turn}& \multicolumn{3}{l}{Rhetorical pieces stayed in the joint document } \\
        \cline{2-4} & Not edited by NNSs & Edited by NNSs & Total \\
        \midrule
         Turn\textbf{\textsubscript{1}} ahead of its next turn\textbf{\textsubscript{2}} taken by an NS&  46 (expected: 53.82)&  13 (expected: 5.18)& 59
\\  
         Turn\textbf{\textsubscript{2}}  after its prior turn\textbf{\textsubscript{1}} taken by an NS&  67 (expected: 65.68)&  5 (expected: 6.32)& 72
\\  
         Turn\textbf{\textsubscript{3}} ahead of its next turn\textbf{\textsubscript{4}} taken by an NS&  71 (expected: 69.32)&  5 (expected: 6.68)& 76
\\ 
         Turn\textbf{\textsubscript{4}}  after its prior turn\textbf{\textsubscript{3}} taken by an NS&  86 (expected: 81.18)&  3 (expected: 7.82)& 89
\\
        \midrule
         Total&  270&  26& 296
\\
    \bottomrule
\end{tabular}
\end{table}
\end{small}

\subsubsection{Comparison between the likelihoods of an NS making edits across different positions.} We performed a similar chi-square test to evaluate the relationship between an NS’s position and edits occurring in the ideational aspect of each rhetorical piece. The variable of position contained four levels: turn\textbf{\textsubscript{1 }}ahead of its next turn\textbf{\textsubscript{2 }}taken by an NNS, turn\textbf{\textsubscript{2 }}after its prior turn\textbf{\textsubscript{1 }}taken by an NNS, turn\textbf{\textsubscript{3 }}ahead of its next turn\textbf{\textsubscript{4 }}taken by an NNS, turn\textbf{\textsubscript{4 }}after its prior turn\textbf{\textsubscript{3 }}taken by an NNS. The variable of edits contained two levels: no vs. yes. 

There was no significant relationship between the two variables: $\chi$2 (3, N= 291) = 6.56, \textit{p} = .09. The likelihood of an NS making ideational edits to each rhetorical piece appeared similar across positions (Table 4). The observed counts for edits were boosted one time when an NS took turn\textbf{\textsubscript{2 }}after an NNS’s turn\textbf{\textsubscript{1}}, but it did not change the significance of the overall results.  

\begin{small}
\begin{table}[h]
    \centering
\caption{NSs’ Edits to the Ideational Aspect of the Rhetorical Pieces }
    \begin{tabular}{p{6cm} p{2.5cm} p{2.5cm} p{0.8cm} }
\toprule
    \multirow{2}{6cm}{NSs’ position in terms of the combination between the order and the turn}& \multicolumn{3}{l}{Rhetorical pieces stayed in the joint document } \\
        \cline{2-4} & Not edited by NSs& Edited by NSs& Total\\
        \midrule
         Turn\textbf{\textsubscript{1}} ahead of its next turn\textbf{\textsubscript{2}} taken by an NNS&  48 (expected: 48.36)&  14 (expected: 13.64)& 62
\\  
         Turn\textbf{\textsubscript{2}}  after its prior turn\textbf{\textsubscript{1}} taken by an NNS&  43 (expected: 49.92)&  21 (expected: 14.08)& 64
\\  
         Turn\textbf{\textsubscript{3}} ahead of its next turn\textbf{\textsubscript{4}} taken by an NNS&  62 (expected: 60.07)&  15 (expected: 16.93)& 77
\\ 
         Turn\textbf{\textsubscript{4}}  after its prior turn\textbf{\textsubscript{3}} taken by an NNS&  74 (expected: 68.65)&  14 (expected: 19.35)& 88
\\
        \midrule
         Total&  227&  64& 291
\\
    \bottomrule
\end{tabular}
\end{table}
\end{small}

\subsubsection{Edits in the format of adding or removing rhetorical pieces across different positions.} The above analyses examined the co-writer’s edits to rhetorical pieces that remained across two adjacent editing turns. The action of adding in new rhetorical pieces or withdrawing old pieces also appeared in our data, but they have not been considered so far. We summarize relevant counts in Table 5. As they indicate, withdrawals happened almost exclusively when a co-writer took the very first turn (i.e., turn\textbf{\textsubscript{1}}) of the entire writing process. Adding new rhetorical pieces was predominantly done by NSs when they took turn\textbf{\textsubscript{2 }}after NNSs’ turn\textbf{\textsubscript{1. }}

\begin{small}
\begin{table}[h]
    \centering
\caption{NNSs’ and NSs’ Edits in the Format of Adding or Removing Rhetorical Pieces}
    \begin{tabular}{p{6cm} p{1.2cm} p{1.6cm} p{0.02cm} p{1.2cm} p{1.6cm}}
\toprule
    \multirow{2}{6cm}{Position in terms of the combination between the order and the turn}& \multicolumn{2}{p{3cm}}{Rhetorical pieces added or withdrawn by NNSs} && \multicolumn{2}{p{3cm}}{Rhetorical pieces added or withdrawn by NSs}\\
        \cline{2-3} \cline{5-6} & Added& Withdrawn & & Added&Withdrawn\\
        \midrule
         Turn\textbf{\textsubscript{1}} ahead of the other co-writer’s turn\textbf{\textsubscript{2}} &  6 &  21 && 9&12\\  
         Turn\textbf{\textsubscript{2}}  after the other co-writer’s turn\textbf{\textsubscript{1}}&  5&  1 && 19&2\\  
         Turn\textbf{\textsubscript{3}} ahead of the other co-writer’s turn\textbf{\textsubscript{4}}&  5&  0 && 8&2\\ 
         Turn\textbf{\textsubscript{4}}  after the other co-writer’s turn\textbf{\textsubscript{3}}&  4&  1 && 9&0\\
         \bottomrule
\end{tabular}
\end{table}
\end{small}

\subsection{Perceived Quality of Task Coordination  }
RQ1 inquired about whether a co-writer’s perceived quality of task coordination would vary with their language background as well as their group’s turn-taking order. We conducted a 2 $\times$ 2 Mixed Model ANOVA to answer this question (Figure 3). The dependent variable was each co-writer’s rating of the perceived coordination quality, as collected via the post-task survey. One independent variable was the co-writer’s language background (language background: NNS or NS). The other was the order of turn-taking (order: turn\textbf{\textsubscript{1 }}taken by an NNS or an NS). Co-writers working on the same joint document were nested in the same group. 

The results indicated that the main effect of the order on the perceived coordination quality was not significant: F [1, 27.9] = 1.22, \textit{p} = .28. The main effect of participants’ language background was not significant either: F [1, 16.3] = 2.77, \textit{p} = .12. However, there was a significant interaction effect between the order and the language background: F [1, 16.3] = 4.60, \textit{p} < .05. 

More specifically, NNSs’ perceived coordination quality remained similar no matter whether they had turn\textbf{\textsubscript{1}} (Mean = 5.90, S.E. = .25) or an NS was at turn\textbf{\textsubscript{1 }}(Mean = 5.98, S.E. = .20; \textit{p} = .81). NSs’ rating of the coordination quality was significantly higher when the turn-taking started with an NNS at turn\textbf{\textsubscript{1 }}(Mean = 5.99, S.E. = .25) than with an NS at turn\textbf{\textsubscript{1 }}(Mean = 5.32, S.E. = .20; \textit{p} < .05). 

\begin{figure}[h]
    \centering
    \includegraphics[width=0.65\linewidth]{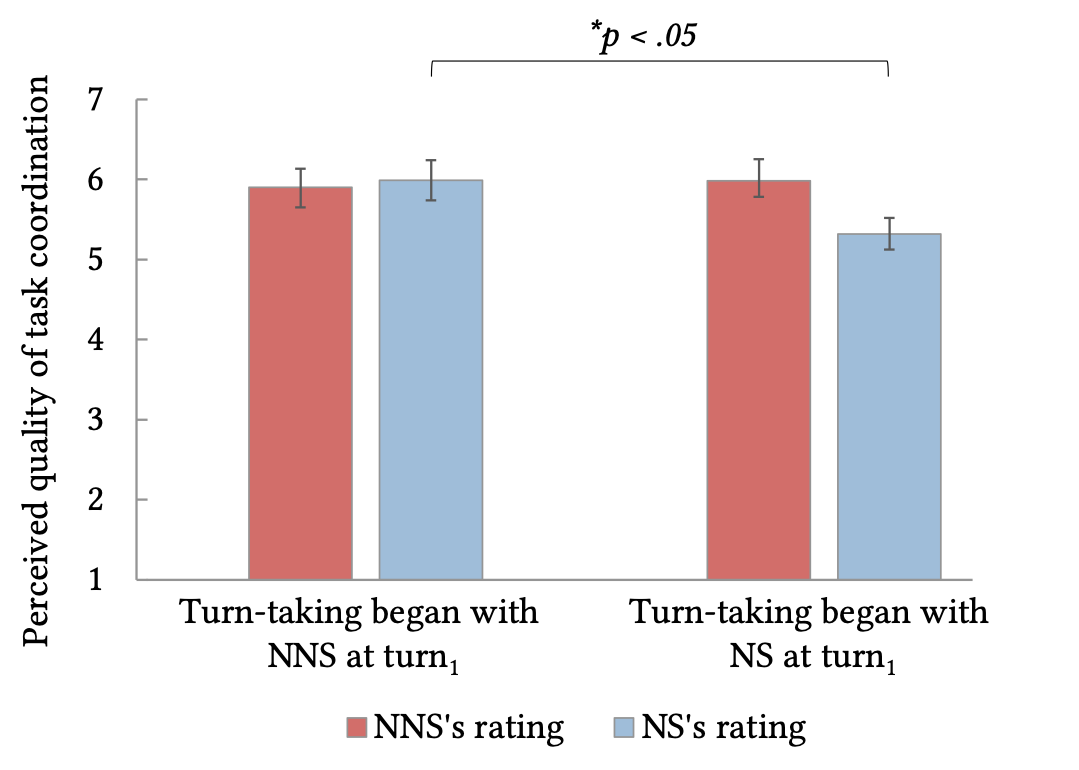}
    \caption{\centering Co-writers’ ratings of the perceived task coordination quality by their individual language background and the order of turn-taking followed by their group.}
\end{figure}

\subsection{Perceived Value of Being Able to Speak English or Japanese }
RQ2 asked whether a co-writers’ perceived value of being able to speak English or Japanese would vary with their language background as well as the order of turn-taking they had followed. This answer was examined through two 2 $\times$ 2 Mixed Model ANOVAs (Figure 4). The dependent variable was each co-writer’s rating of their perceived value of having language ability in English or Japanese. One independent variable was the co-writer’s language background (language background: NNS or NS). The other was the order of turn-taking (order: turn\textbf{\textsubscript{1 }}taken by an NNS or an NS). Co-writers working on the same joint document were nested in the same group.

The results indicated that the main effect of the order on the perceived value of being able to speak English was not significant: F [1, 25.8] = .02, \textit{p} = .89. There was a significant main effect of language background: F [1, 30.2] = 30.41, \textit{p} < .01. That is, NSs of English (Mean = 6.80, S.E. = .18) perceived their language ability in English to be more valuable to the current task context than NNSs (Mean = 5.32, S.E. = .18; \textit{p} < .01). However, the interaction effect between the order and the language background was not significant: F [1, 30.2] = .37, \textit{p} = .55.

Meanwhile, the main effect of the order on the perceived value of being able to speak Japanese was not significant: F [1, 27.5] = 1.24, \textit{p }= .27. There was a significant main effect of language background: F [1, 29.5] = 7.29, \textit{p} = .01. The interaction effect between the order and the language background was also significant: F [1, 29.5] = 7.05, \textit{p} = .01. More specifically, NNSs of English perceived their language ability in Japanese to be more valuable when the turn-taking started with themselves at turn\textbf{\textsubscript{1}} (Mean = 5.25, S.E. = .42) than with NSs at turn\textbf{\textsubscript{1 }}(Mean = 3.60, S.E. = .45; \textit{p} = .01). NSs’ perceived (hypothetical) value of them being able to speak Japanese appeared similar no matter if the order of turn-taking started with themselves at turn\textbf{\textsubscript{1 }}(Mean = 3.58, S.E. = .45) or with NNSs at turn\textbf{\textsubscript{1 }}(Mean = 2.89, S.E. = .42; \textit{p} = .26).

\begin{figure}[h]
    \centering
    \includegraphics[width=1\linewidth]{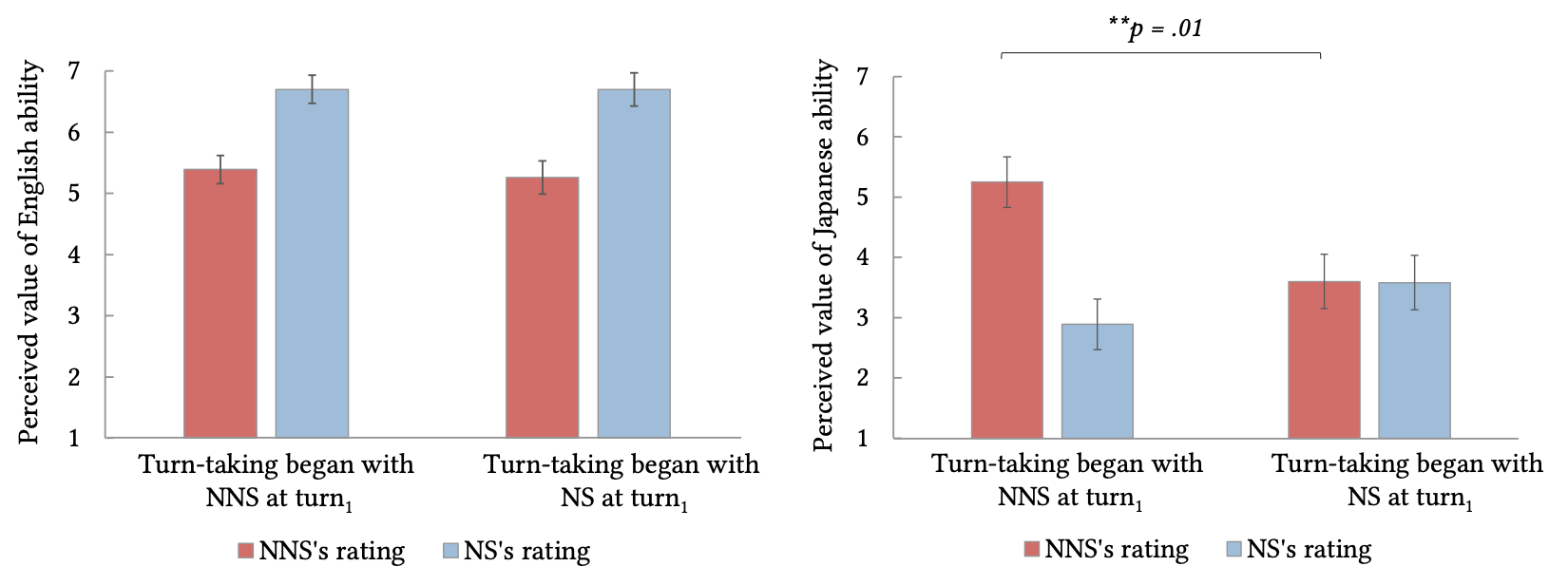}
    \caption{\centering Co-writers’ ratings of the perceived value of English (left) or Japanese (right) ability by their individual language background and the order of turn-taking followed by their group.}
\end{figure}

\section{QUALITATIVE FINDINGS}

We performed a thematic analysis to extract insights from post-task interview responses [11]. This analysis was led by three researchers on our team, given their experience working with both English and Japanese speaking participants. At the beginning of the process, researchers read through all the interview transcripts and familiarized themselves with the data. After that, each person coded an exclusive subset of the data, then reviewed and refined their codes at a group session. Through multiple iterations between the individual and collective sessions, we arrived at a final list of codes and themes. These codes and themes described the co-writers’ perceived challenges and benefits of writing with the other person (RQ1, RQ2), as well as the value of using AI-powered editing tools (RQ3). Some of the perceptions appeared sensitive to the order of turn-taking between co-writers, while others did not. We detail our findings below. 

\subsection{Task Coordination as Co-Writer Groups} 

\subsubsection{NNSs’ perspectives separating ideation from expression.} NNSs participating in this research commented extensively on their struggles with expressing ideas in English. Many described themselves as “\textit{not good at articulating them in a second language}” and “\textit{[experiencing a] loss of control over the nuances}.” Associated with this perception, most NNSs aligned their task contribution towards “\textit{diversifying the perspectives communicated in the article}” and “\textit{enriching the co-writer’s initial ideas}.” Participants felt that they devoted significant effort to the ideational aspect of the collaborative writing, although their endeavors often did not result in a substantial volume of text going into the document: 

\begin{itemize}
    \item[] \textit{“My co-writer and I worked on an article in response to the reader’s question about education and the use of digital devices. My co-writer offered ideas and examples from a U.S. perspective, while I presented some additional information in the contexts of Japan and Asia in general. I believe my contribution was valuable because it expanded upon what we had as a group. }” [NNS-22; NNS at turn\textbf{\textsubscript{1}}]
\end{itemize}

\begin{itemize}[]
    \item[] \textit{“When I received the document back from my co-writer, I found that they used some compelling examples to support our ideas. Some of the examples could be strengthened with further details. So, I went online to find relevant information and incorporate it into the text. I didn’t write very long, and the language was probably not good. But the co-writer later edited the content to make improvements.}” [NNS-9; NS at turn\textbf{\textsubscript{1}}]
\end{itemize}

\subsubsection{NSs’ perspectives considering content production as a whole.} In contrast to NNSs’ points of view, NSs perceived the English writing task to be “\textit{easy to manage.}” Participants noted that the document usually “\textit{looked quite good}” after they had performed one full round of editing. By the third or fourth turn, several NSs had shifted their focus towards refining minor details. Some NSs reported that the ideas of NNSs inspired their own writing in subsequent turns. Nevertheless, they all noticed that the volume of NNSs’ content production was, in general, small: 

\begin{itemize}
    \item[] \textit{“In their first section [turn}\textbf{\textit{\textsubscript{1}}}\textit{], my co-writer included most of my content in the writing. They didn’t generate a ton of [additional] content. However, I noticed that they made some really good points that I had never brought up. When it was my turn [turn}\textbf{\textit{\textsubscript{2}}}\textit{], I made sure to incorporate those points in a concise way. }” [NS-25; NNS at turn\textbf{\textsubscript{1}}]
\end{itemize}

\begin{itemize}
    \item[] \textit{“I incorporated both people’s [content] into the writing during my first section [turn}\textbf{\textit{\textsubscript{1}}}\textit{], and then in my second section [turn}\textbf{\textit{\textsubscript{3}}}\textit{]. But, from my memory of the collaboration, I don’t feel there was a lot that they did. Especially in the latter half of the process, I felt like I was doing much more than they were.}” [NS-6; NS at turn\textbf{\textsubscript{1}}]
\end{itemize}

\subsubsection{Coordination challenges when the turn-taking started with NSs.} Overall, all participants reported more negative coordination experiences when the turn-taking began with an NS at turn\textbf{\textsubscript{1}}, as opposed to having an NS act following an NNS’s turn\textbf{\textsubscript{1}}. For NNSs, this turn-taking order increased their difficulty in contributing to the joint document without “\textit{mess(ing) up what the NS had already completed.}” Many NNSs spent considerable time scrutinizing the document as well as relevant Japanese and English information sources, attempting to identify areas where additional edits might be needed. Yet, they often refrained from making those edits:

\begin{itemize}
    \item[] \textit{“My co-writer drafted the entire article. I played a supporting role by inserting the ideas I could think of into their writing. However, it wasn’t always possible to integrate my content. After my co-writer’s first turn [turn}\textbf{\textit{\textsubscript{1}}}\textit{], I saw that they had put together an impressive piece. I searched for relevant materials during my turn [turn}\textbf{\textit{\textsubscript{2}}}\textit{], but I wasn’t sure about the right places to integrate them.}” [NNS-4; NS at turn\textbf{\textsubscript{1}}]
\end{itemize}

\begin{itemize}
    \item[] \textit{“There was little I could do to help after reading the article edited by my co-writer. It felt like they had finished the whole thing. I tried very hard to make further improvements. During my turn of editing, I read through the article multiple times. I didn’t want to update the content unless it was really necessary.}” [NNS-6; NS at turn\textbf{\textsubscript{1}}]
\end{itemize}

The above concerns resulted in NNSs performing far fewer edits than NSs anticipated. All NSs taking turn\textbf{\textsubscript{1 }}recalled that, when they revisited the document at turn\textbf{\textsubscript{3}}, they were surprised by how little it had changed. NS-6 shared her frustration, which represents the experiences of other NSs: 

\begin{itemize}
    \item[] \textit{“I felt a bit disappointed. I was hoping for ideas that weren’t just mine. I knew they had some good points, but I was hoping there could be more. As I think of my expectations, yes, it was a lot easier for me to write extensively because I was using my native language. I wasn’t overly upset [with my co-writer]. But we were working on a collaborative task. I was interested in what they would have to say on the topic.}” [NS-6; NS at turn\textbf{\textsubscript{1}}]
\end{itemize}

\subsection{Use of AI-powered Editing Tools and Its (Unexpected) Effects}

\subsubsection{NNSs’ use of AI-powered editing tools for translanguaging.} From the video clips referred to during the interview sessions, we observed that all NNSs had utilized at least one machine translation system (e.g., Google Translate, DeepL) for information processing between Japanese and English. Some also turned to paraphrasing tools (e.g., Wordtune, Ginger) for alternative expressions of their English sentences. The use of AI-powered tools enabled NNSs to perform “translanguaging [85]” or the use of their multilingual repertoire to a certain extent; however, it did not place NNSs on equal footing with NSs. As NNSs noted, the text generated by AI-powered tools often appeared “\textit{natural sounding}” and “\textit{authentic.” }They had to devote significant effort assessing the pragmatic appropriateness of the tool’s outputs. We detail two instances below.

NNS-21 reflected on an instance where she turned to a paraphrasing tool, Ginger, to improve the English sentences she had written. In one section of the article, this participant sought to express empathy with the reader. She composed the following English sentences: “\textit{I feel the same way too. I know that when I feel like that, I think I want to do it or I can do it}.” She copied her sentences into Ginger and reviewed the tool’s outputs for a long while. During the interview session, NNS-21 talked about her thinking process while assessing the tool’s outputs: 

\begin{itemize}
    \item[] \textit{“I often use paraphrasing tools to explore different ways of writing the same sentence. They can be helpful especially when I want the sentence to be longer, contain a wider variety of vocabulary, or to have a more formal or informal tone. In this example, I checked Ginger for insights but couldn’t tell the exact difference among those outputs. To move forward with the writing, I just used my initial sentence with the hope that my co-writer would make revisions if something went wrong.}” [NNS-21; NNS at turn\textbf{\textsubscript{1}}]
\end{itemize}

NNS-11 walked us through a different instance where he sought to find the English translation for his initial writing in Japanese, which would literally translate to “\textit{I have come to like myself}.” To identify the most appropriate English expression, this participant iterated multiple times between adjusting his Japanese inputs and reviewing the English outputs from translation tools (Appendix C). The corresponding video clip revealed that it took him around 5 minutes to come up with the final sentence, which consisted of only a few words. A similar process was also found among other NNSs according to self-reports. After showing us the video clip, NNS-11 added:

\begin{itemize}
    \item[] \textit{“I have to admit that I did not arrive at the most satisfying answer. This final sentence should be close to what I wanted, but I wasn’t entirely sure. It took me a long time to craft this sentence. I thought I’d better stop there.}” [NNS-11; NS at turn\textbf{\textsubscript{1}}]
\end{itemize}

\subsubsection{NSs’ use of AI-powered editing tools for proofreading.} NSs in this research reported minimal use of AI-powered tools over the entire task process, which was confirmed by their video recordings. About half of these participants employed Google Docs’ built-in spelling and grammar check function to spot issues with their writing. Others leveraged Grammarly to proofread their articles. However, more advanced functions, such as automated paraphrasing or text generation, were not considered by these participants. This usage pattern stemmed from NSs’ confidence in their ability to tailor the writing to its social context and audience. As stated by NS-31:

\begin{itemize}
    \item[] “\textit{As a human, I am able to get a holistic view on what this writing is meant to be for the reader. AI tools will do a satisfactory job at the surface level, for example, to switch some words up or to make something more formal. However, they do not interpret [the social context of] the prompt as I do. If I have to get something more than proofreading, I feel a human reader’s advice will be more helpful}.” [NS-31; NNS at turn\textbf{\textsubscript{1}}]
\end{itemize}

Moreover, many NSs characterized writing as an activity to demonstrate the content producer’s “\textit{voice}” and “\textit{autonomy}.” AI-powered tools today are capable to generate humanlike text. However, what human writers value most are the choices they make to convey their unique intentions: 

\begin{itemize}
    \item[] “\textit{If I fed the prompt into an AI tool and it returned a flawless article, I wouldn't feel like ‘wow, it did such a fantastic job for me.’ It’s not just about having the AI comprehend instructions and generate information. The point of writing is the communication between the writer and the reader, and that’s what truly matters.” }[NS-25; NNS at turn\textbf{\textsubscript{1}}]
\end{itemize}

\begin{itemize}
    \item[] “\textit{I have the flavor of what I'm trying to say. I need the freedom to make my own choices [about language use]. If I've used the word ‘concern’ earlier in the paragraph, I may want to avoid}\textit{ repeating it again. Or, if I'm trying to imply a deep concern rather than a general worry, I will}\textit{ pick words with different connotations. Grammarly and other tools can suggest a word, but, very likely, I will just reject their suggestions}.” [NS-8; NS at turn\textbf{\textsubscript{1}}]
\end{itemize}

\subsubsection{Disruptions to the interpersonal dynamics between co-writers.} We noticed that NNSs’ use of AI-powered tools disrupted the interpersonal dynamics between co-writers in unexpected ways. Specifically, all NSs reported that, prior to the formal start of the collaborative work, they were “\textit{prepared to write more}” to accommodate their NNS co-writers. However, the actual writing delivered by NNSs sometimes left NSs confused about their co-writer’s level of English proficiency. Most NSs did not realize their co-writer had made extensive use of translation and/or paraphrasing tools to produce English content. Nor did they have much awareness of NNSs’ efforts in assessing the tool’s outputs. As a result, many NSs misperceived their co-writers as skillful English writers who chose to act passively in the collaborative work:

\begin{itemize}
    \item[] “\textit{At the beginning, I didn’t expect the other person to write a lot}. \textit{English was not their mother tongue. So, I thought I would probably act more like a leader. Later on, I read their writing. It left me the impression that they actually had good grasp of the English language. I ended up wondering why they had generated so little content if they were able to write in good English.}” [NS-3; NS at turn\textbf{\textsubscript{1}}]
\end{itemize}

Furthermore, we witnessed multiple cases in which NSs misinterpreted the extent of NNSs’ agency in their English writing. To detail two examples, NNS-1 and NNS-26 both began part of their writing in Japanese and with a soft tone. They then used Google Translate to generate the English version, hoping their co-writers (i.e., NS-1 and NS-26) could revise the English later. With no awareness of the above process, the NS co-writers in both cases decided to leave the writing as it was, although the content in fact appeared problematic to them:

\begin{itemize}
    \item[] “\textit{Here, my co-writer said, 'you may be on the verge of becoming so dependent on social networking sites.’ It was almost blaming [the reader]. I didn’t want the reader to feel like ‘oh my God, something is wrong with me,’ but I didn't rewrite the sentence. I didn't want them to feel like I was taking over the piece}.” [NS-26; NNS at turn\textbf{\textsubscript{1}}]
\end{itemize}

\begin{itemize}
    \item[] “\textit{My co-writer had a sentence, ‘why don’t you stop comparing yourself to others and focus on yourself?’ I thought it carried too much of an accusatory tone. We should have a more polite way of making suggestions to the reader. However, I wasn’t sure if I should make a direct change [because] it was different from [changing] something grammatical. I didn’t want my co-writer to feel I was judging their intention.}” [NS-1; NS at turn\textbf{\textsubscript{1}}]
\end{itemize}

\section{DISCUSSION}

To recap, our work demonstrates that NNSs have the potential to make unique contributions to the ideational aspect of collaborative writing. That said, their potential tends to be suppressed when NNSs act as the late-mover in coordination with NSs. AI-powered editing tools facilitate NNSs’ content production by enabling translanguaging; however, these tools often require the user’s close monitoring to ensure output quality, introducing new challenges to an NNS and the NS-NNS coordination. Moreover, our findings indicate the importance of providing each co-writer dedicated and protected individual access to the joint document, especially when their group coordination takes place in a multilingual setting. The workflows or turn-taking orders adopted in the current research acknowledge each party’s full rights and responsibilities over the entire document. This approach differs from some previous ones that emphasize more fine-grained divisions regarding who holds the right to control which part(s) of the writing (e.g., [52]). Below, we discuss our work’s implications for successful collaboration between NSs and NNSs (Section 6.1), deliberate use of AI-powered tools by human co-writers (Section 6.2), and system design that promotes the linguistic inclusion and equity in collaborative writing (Section 6.3).

\subsection{Coordinating Expression and Ideation as a Co-Writer Group} 

Our work is not the first piece to discuss the role of workflow or turn-taking order in collaborative writing. However, prior work in this space has rarely examined coordination between co-writers with diverse language backgrounds. As a result, the unique position of a late-mover was usually considered in terms of their needs to recognize open opportunities for new edits (e.g., [6, 70]), to carefully interpret the current flow of the document (e.g., [76, 88]), and to fit their writing to the narrative established by preceding edits (e.g., [56]). Several interview studies have found that the above needs often promote a late-mover’s reflective thinking about their edits, but not hinder their contribution to the writing [8, 50, 53]. 

The current research complements existing literature by demonstrating that a co-writer’s interpretation of open opportunities to make edits is likely to vary according to their working language proficiency. NNSs in this study identified many fewer opportunities than NNs to edit the expressional aspect of their joint document. As revealed in our data, little of this contrast was rooted in a participant’s ability to recognize and correct grammar errors in English writing. Instead, what mattered was the subtlety of language use, such as different wordings and tones that convey the similar idea. NSs tweaked the document’s wording throughout the entire writing process, whereas NNSs’ edits to this expressional aspect always remained limited. 

Notably, the current results concerning expressional edits does not support our prediction that was derived from existing literature. In that literature, the likelihood of an NNS editing the expressional aspect of an article was supposed to be enhanced after they had reviewed an NS’s edits. We suspect that the absence of such enhancement in our study is due to the context of collaborative, rather than individual, writing. Specifically, prior work on individual writing positioned NNSs as the solo custodian for the content production (e.g., [41, 42]). NSs in that context gave feedback on NNSs' writing, but they did not share responsibilities of task performance as a co-writer. In our task context, NNSs and NSs both acted as part of a co-writer group. NSs’ full proficiency in the working language positioned them to lead the edits made to the expressional aspect of the document. Our data does not provide direct evidence regarding whether NNSs’ ability to make expressional edits varied according to the order of turn-taking between NNSs and NSs. Rather, it reveals NNSs' intended choices to hold back on introducing changes to English expressions used in the document. 

More importantly, NNSs in our study performed less ideational edits when they acted after their co-writers, as opposed to ahead of their co-writers. This finding underscores the importance of considering the timing of NSs’ edits in relation to that of NNSs. From the analysis of post-task reflections, we learned that it was natural for NSs to weave their thoughts into the existing content and, simultaneously, polish the language of the entire piece. While this fashion of editing appeared effective for NSs, it raised the bar for NNSs to either expand on the ideas already discussed in the document or introduce new ones. NNSs ended up perceiving the document as being too ready to be edited from both expressional and ideational aspects. 

For the realization of NNSs’ full potential, our research suggests that NNSs demand the space to experiment with temporary and often unrefined ways to articulate ideas. We identified turn-taking order as a critical condition that either opens or suppresses that space.

\subsection{Working with AI-Powered Editing Tools Versus Human Co-Writers}

Participants in our research discussed three levels of edits that they ever considered performing. One level concerned the grammar rules of English writing. The second was related to style, as revealed in word choices and sentence tone. The third focused on the ideas conveyed through the sentences. The distinctions among these types of edits are encapsulated in Kraut et al.’s discussion of equivocality, or the extent to which alternative solutions can be applied during the editing of a joint document [50]. One question worth asking is: between AI-powered editing tools and human co-writers, who is considered more suitable to manage the edits at each level, and why? Our research findings offer important insights into this question. 

Specifically, neither NSs nor NNSs in our participant pool elaborated much on their thinking process of making grammar edits. Aside from the fact that all participants were able to manage basic English grammar, the low equivocality of grammar rules might also lead people to perform such edits without additional concerns. NNSs were open to having the syntax of their writing revised by the NS co-writer or automated proofreading tools (e.g., Grammarly), depending on the convenience of access to each. NSs leveraged similar tools for efficiency. 

In contrast, participants perceived high equivocality when tweaking the style of an expression as well as the nuances of meaning that expression could convey. NSs and NNSs both remained vigilant about allowing AI-powered tools to make such edits on their behalf; however, they developed different practices. For NSs, their high English proficiency enabled them to act without a tool’s assistance; they deliberately avoided using AI-powered tools to manage the nuances of their writing. They also prioritized maintaining the original tone and intent of the NNSs’ writing, even though they could have offered alternative phrasings. For NNSs, AI-powered translators and paraphrasers enabled the discovery of English expressions that would otherwise have remained unknown to them. Our NNS participants frequently turned to these tools for editing suggestions. That said, they strived to assess the propriety of the tool’s output rather than blindly adopting it. These findings indicate that, when it comes to edits involving high equivocality, human co-writers value their agency displayed in the content production more than convenience or efficiency. AI-powered tools, ideally, should support the preservation and negotiation of this agency. 

Furthermore, our work suggests the involvement of AI-powered tools can negatively affect the interpersonal dynamics between co-writers. This finding contrasts with those reported by prior scholars in important ways. In particular, early CSCW studies on NNSs’ use of language processing tools, such as machine translation, often considered the tool as a source of errors. They found that NSs were usually not good at differentiating machine-generated errors from disfluencies produced by an NNS (e.g., [26]). From there, they argued using tools would produce a positive effect because NSs might attribute communication errors to the tool rather than to NNSs. Participants in our research rarely discussed the errors or disfluencies found in the tool’s outputs. Conversely, it was the high-fidelity of those outputs against human language that posed challenges to NNSs and to the group’s coordination. A similar observation has also been reported in a small number of recent research on multilingual conversations (e.g., [72]). 

As the performance of AI-powered tools continues to advance, we urge more CSCW research to consider its multifaceted effects on human collaborative work. The current research was conducted before the widespread adoption of ChatGPT and other LLM-based language processing tools by the general public. Nonetheless, our findings indicate several opportunities and pitfalls to be mindful of in the ongoing development of such tools catering to a diverse user base. For NNSs, in particular, one promise of advanced AI tools lies in their capacity to tailor responses to user-generated requests. Recent literature has demonstrated instances in which human writers leverage these tools for text editing based on user-defined aims (e.g., [43]), generating narrative elements in a given context (e.g., [93]), and integrating user-specified vocabularies or themes into the text output (e.g., [14]). NNSs are likely to perceive a greater sense of agency in the interaction with advanced AI tools, as they no longer have to retrospectively assess the discrepancies between original text and its edited versions. The conversational interface adopted by these tools consolidates the above potential by lowering NNSs’ barriers in specifying their intent. 

Yet, the increasingly natural interaction between advanced AI tools and their users can obscure problems with the tool’s output. NNSs in our research tended to equate humanlike text outputs with high-quality outputs. NSs expressed much stronger value of text written by NNSs than that generated by tools; however, they encountered significant difficulties in distinguishing the actual source of the text. When it comes to more recent tools, such as those built upon LLM, users will be required to go through even more sophisticated processes to recognize inaccurate or uncredited text outputs [39]. Emerging discussions surrounding hallucination (e.g., [4, 40]) and the low reliability in text source detection (e.g., [68, 73]) have provided support for this conjecture. 

Returning to the focus of our work, we are concerned that the use of ChatGPT and other LLM-based tools may lead to a situation where the rich get richer and the poor get poorer. That is, people who possess existing skills and knowledge in producing the target content can better leverage AI-powered tools for productivity. Those who lack the necessary skills and knowledge will find themselves at a persistent disadvantage, similar to what we have observed with NNSs in the current research. 

\subsection{System Design for the Linguistic Inclusion and Equity in Collaborative Writing}

\subsubsection{Scaffolding the coordination of expression and ideation via separated steps.}The current research suggests that NNSs with limited working language proficiency often struggle to manage multiple aspects of English writing simultaneously. When NSs perform their editing pass at an early turn, NNSs may perceive a limited space to add ideas using unrefined expressions. These findings imply that separating the expression and ideational aspects of the coordination may benefit co-writer groups whose members have diverse language backgrounds. 

In the broader HCI literature, many studies have adopted this divide-and-conquer approach to facilitate the participation of disadvantaged individuals in collaborative work. For example, Li et al. designed a conversational agent for multiparty conferencing involving NNSs and NSs of English. The agent was programed to identify speaking opportunities for NNSs. It relieved NNSs from the burden of finding suitable gaps to chime in during an ongoing conversation [54]. Das et al. studied the collaborative writing experience of people with vision impairments. Their findings suggested that separating the audio signals of the document’s initial content and suggested edits would help participants manage their task more effectively [18, 19]. 

Moving to the context of collaborative writing between NNSs and NSs, we envision that future systems could remind both parties to focus on the ideational aspect of their joint work during earlier stages of group coordination. If the system senses that a co-writer is spending considerable time refining the expression of a sentence, it could prompt this person to refocus attention on the clarity of their sentence rather than on full subtleties. The system may also assign auto-generated labels to such sentences, indicating the need for additional edits at a later stage. 

\subsubsection{Raising awareness of NNSs’ contributions across various aspects.} NNSs’ contributions to the collaborative writing in our study has been evaluated by various measures. Some are more sensitive to NNSs’ potential than others. Regardless of the order of turn-taking, NNSs consistently contributed less than NSs in terms of lexical changes introduced to the joint document. They also displayed low confidence and were unlikely to edit the expressional aspect of the document. However, the unique potential of NNSs lies in their ability to introduce complementary ideas and/or elaborate on existing ideas with supplementary information. This contribution is likely to be overlooked if we consider lexical or expressional edits as the primary measures. 

Previous HCI research has developed a rich set of tools that assist in co-writers’ awareness and evaluation of one another’s contributions. Unfortunately, few of them have equitably considered people who write in their non-native language with limited proficiency. One group of existing tools, such as Time Curve [2], depict the temporal evolution of the document’s lexical content as a whole. These tools aid co-writers in tracking the stagnation and oscillation of their overall editing process, but they do not specify the contribution made by each person. A second group of tools, including HistoryFlow [86] and DocuViz [87], adopt the format of a Sankey diagram to represent content additions, deletions, and moves made by each co-writer. However, they fail to capture the function of those edits in terms of their relation to the expression and/or ideational aspect of the document content. The last group of tools enable co-writers to play back video recordings of their writing process (e.g., [13, 83]). While this method provides a person’s full editing history, the information is presented with too many details to distill insights. 

Our research emphasizes the importance of acknowledging and identifying NNSs’ contributions to the ideational aspect of a document. To this end, we propose that future systems for collaborative writing should support the monitoring of a document’s content changes across multiple levels. When people attempt to comprehend each co-writer’s contributions as either a participant of the work or a third-party evaluator, they can leverage the Sankey diagram given by HistoryFlow or DocuViz for an overview. On top of that, more detailed information can be embedded into each segment of the Sankey diagram. This information should specify how the selected edits relate to the expression and/or ideational aspects of the writing, as well as who has participated in making those edits. 

\subsubsection{Making the use of AI-powered editing tools salient.} Last but not least, the current research suggests that future collaborative writing systems should pay special attention to content produced by AI-powered editing tools or with the tool’s assistance. NSs are likely to hold different assumptions about their co-writer’s writing proficiency, as well as the types of edits to coordinate, depending on whether they have realized NNSs’ interaction with AI-powered tools. Thus, it can be beneficial to provide NNSs and NSs with equal awareness of each party’s tool usage. 

We believe it is feasible for future systems to trace a co-writer’s use of AI-powered tools. For instance, by mining computer logs generated over the task period, the system will be able to identify places in a document where the co-writer paused their writing and turned to machine translators or paraphrasers for editing suggestions. In the case of more recent tools, such as ChatGPT, the system can store and retrieve a user’s full interaction history with their consent. 

That said, we caution against implementing such functions without further research. Previous studies in the context of teamwork have documented instances where NNSs hid their use of language processing tools from NS colleagues for impression management (e.g., [25]). Comparing that research with our current study, we suspect that the impulsive disclosure or concealment of NNSs’ use of tools could both negatively affect the coordination between NNSs and NSs. Future research should investigate ways to make the use of AI-powered tools salient, but at a level that is appropriate for those using them. 

\section{LIMITATIONS AND FUTURE DIRECTIONS }

The methodological choices of our study design allowed us to examine how collaborative writing between NNSs and NSs was affected by variables of interest. That said, the advantages of our approach come with their own limitations. We carefully reflect on the ecological validity of this research against the full spectrum of collaborative writing practices involving various possible workflows (Section 7.1), more than one way to consider a person’s language use ability (Section 7.2), and the often-intertwined effects of language and culture (Section 7.3). At the end of each subsection, we outline directions for future work that can build upon and complement ours. 

\subsection{Manipulation of the Turn-Taking Order}

There is a strong theoretical underpinning for our positioning of turn-taking order as the primary variable of interest. Prior HCI and CSCW research has emphasized the importance of turn-taking order in shaping co-writer groups’ coordination behavior; however, it rarely considered NNSs’ potentials and struggles during this process. The core contributions of our work include the deductive building of relevant hypotheses against literature pertaining to NNSs’ individual writing and, more importantly, offering empirical evidence to show that those hypotheses do not all hold true. As argued by other CSCW scholars performing hypothesis-testing work, the nature of such work requires researchers to specify the variables of interest, as well as the levels to be considered for each variable, in order to study them rigorously and effectively [28]. 

But what about other possible levels of the core variable or additional variables that are not covered in the study design? In the case of our current research, why didn’t we prioritize other formats of collaborative writing, such as simultaneous work or unstructured turn-taking between co-writers? We acknowledge these are important questions for assessing the eco-validity of our research, and we offer further reflections in this regard. 

First of all, structured turn-taking has been identified as a workflow frequently adopted by co-writer groups in the real-world (e.g., [56, 70, 76]), although it is not the only possible workflow (e.g., [47, 66, 79]). In HCI and CSCW, substantial work has described the reasons motivating co-writers to opt for structured exchanges of editing turns (e.g., [1, 8, 10, 52, 58, 70, 88]). Much of this work was performed after the technical challenges of document change tracking and version control had already been resolved. In particular, Boellstorff et al. offered an autoethnographic reflection on their experiences with collaborative writing under different workflows. They noted that structured turn-taking helped minimize “process loss,” such as the mental and physical burdens of synthesizing parallel changes to the same content, and the coordination delay caused by lack of clarity about the content’s ownership [10]. Wang et al. interviewed individuals with extensive experience in collaborative writing. Participants reported a pronounced preference for avoiding simultaneous writing, as it often led to feelings of intrusion into each co-writer’s private thought space [88]. Andr\'e et al. presented experiment results demonstrating that, as opposed to simultaneous work, structured turn-taking better protected co-writer’s sense of territoriality and promoted editing of one another’s writing [1]. 

We believe the above value and preference of structured turn-taking extends well from a general setting, where the co-writer group’s linguistic composition is unspecified, to the case of NS-NNS collaborative writing. Not only that, the unique dynamics of NS-NNS collaboration often make structured turn-taking a necessary condition to enable NNSs’ meaningful participation in the joint work, if at all. Previous CSCW research on NS-NNS meetings has repeatedly highlighted this phenomenon. As one example, Yamashita et al. found that a flexible exchange of speech turns often resulted in NNSs remaining silent during remote meetings with NSs. To address this issue, their research explored the technical solution of using artificial gaps to force the opening of speech opportunities for NNSs [91]. A more recent project by Li et al. considered the same challenge to NS-NNS meetings and leveraged an automatic agent to help secure NNSs’ turns to speak [54]. When it comes to NS-NNS collaborative writing, we were concerned that a flexible exchange of editing turns would, similarly, create an environment that is too unfriendly to elicit NNSs’ contributions. Indeed, even in our research where NNSs can already secure writing opportunities through structured turn-taking, the data still demonstrated a low level of NNSs’ lexical editing in the joint document across turns (e.g., Figure 2). Therefore, it is arguable that, if choosing the option of flexible turn-taking, we might not be able to obtain sufficient data for the investigation of NNSs’ contributions. The choice of structured turn-taking provided us the space to identify a workflow that can better realize NNSs’ potential in collaborative writing with NSs. 

Notably, as one of the first studies to examine NS-NNS collaborative writing, our work did not exhaust all possible configurations of NS-NNS co-writer groups or the full variety of their workflows. Real-world collaborative writing could involve multiple NSs and NNSs. In such settings, it would be challenging to pinpoint the dynamics between co-writers through hypothesis-testing research, as was done in our current work. Also, although the exchange of editing turns can usually be decomposed into segments that began with one party or another, an increase in the number of co-writers will inevitably complicate the choice of workflow. Having limited means for comprehensive tracking of a person’s writing behaviors can add even more complexity to the study of workflow at the individual and group levels. In our work, for example, the editing histories and each person’s self-obtained videos suggested that participants did not engage in writing behaviors outside of their editing turns; yet there could be a possibility that additional exchanges did occur between co-writers and somehow fell through the crack of our data collection protocol. Future scholars should continue to investigate NS-NNS collaborative writing in various settings, leveraging methods that can best suit the specific research focus and needs of their context. Our current research provides one benchmark for such investigations. 

\subsection{Language Use Ability and Its Multifacetedness}

Our current work is grounded in cross-disciplinary literature, including rich contributions from HCI and CSCW scholars (e.g., [12, 17, 44, 91, 95]), within a multilingual setting. This literature usually conceptualizes a person’s language use ability as part of their demographic background. It studies the interactions and/or comparisons between individuals who speak different native languages and/or have imbalanced fluency in one designated working language. Following this tradition, we examined collaborative writing between NSs of English and NNSs who were able to produce content in English but with limited proficiency. 

In the broad literature on collaborative writing, the impact of language use ability on a co-writer group’s task performance has been approached from more than one way. For instance, participants in several interview studies have reported that, when acting as the low-expertise person in a collaborative writing project, they lacked the confidence of language use or literacy in the given topic domain. As a result, they often refrained from editing text produced by those claiming high expertise (e.g., [8, 52, 88]). These findings imply similar dynamics across settings where co-writers possess different levels of competence in language use, regardless of whether this competence stems from people’s language background or their domain expertise. Thus, we suspect that the late-mover disadvantage observed with NNSs in multilingual collaborative writing may also constrain individuals who act as low-expertise members of co-writer groups. 

While we were aware of the expertise aspect of language ability, we chose to let participants work on topics familiar to both NS and NNS members of the same co-writer group. This setup minimized the confounding effects of domain expertise on our participants’ task performance. Future research should consider exploring the intersection of these research lines. For instance, real-world scenarios of collaborative writing may involve NSs working with NNSs who possess greater domain expertise than themselves. This raises several interesting empirical questions, such as whether NNSs can leverage their expertise to compensate for less-than-perfect proficiency, and how the workflow between co-writers with different language backgrounds and domain expertise can be optimally structured to maximize the potential of both parties. 

\subsection{Culture as an Underexplored Factor} 

One notion often intertwined with a person’s language ability is their cultural background. In the current research, our data revealed clear associations between the late-mover disadvantage experienced by NNSs and their limited proficiency in English as a second language. Nevertheless, we wondered if the data could shed further insights on collaborative writing when we shifted our perspective from NS-NNS coordination to an intercultural one. Below, we detail a couple of takeaways derived from this thought experiment. 

Collaborative writing holds the promise of harnessing diverse thoughts and perspectives from multiple minds. When co-writers differ in their cultural backgrounds, there is an amplified chance for each party to introduce exclusive information from outside another’s daily information world. In the context of our research, NNSs’ unique contributions to the ideational aspect of the joint document were partially rooted in their life experience in Japan. One fundamental building block of this experience is their ability to gather and process information in the Japanese language, which, by itself, does not disadvantage our participants but rather benefits them. 

Besides shaping the information to be communicated, a person’s cultural background can also influence their style of language use. Previous research has found that Japanese people usually favor indirect communication as well as the utilization of multiple clues to convey an integrated meaning (i.e., high-context communication style), whereas North Americans prefer the opposite (i.e., low-context communication style) [30, 31, 46, 90]. This cultural difference may exacerbate the difficulties NNSs face in managing their pragmatic use of the English language. As demonstrated by our data, NNSs struggled to assess the equivalence between their Japanese sentences and the English translations, even when they had no problem comprehending the semantic meaning of the latter. They also hesitated to alter the narrative style already set up by NSs in preceding turns. 

Given the above reflections, we believe the late-mover disadvantage experienced by our NNS participants is also observable in collaborative writing involving NNSs who are not Japanese but also have limited capacity in mastering the working language. The more NSs and NNSs differ in their culture-based communication styles, the greater constraint NNSs would face in their attempts to contribute as co-writers. This second claim requires verification by future studies. 

\section{CONCLUSION}

Collaborative writing involving individuals from diverse language backgrounds has received little attention in CSCW literature. The current paper presents our empirical research that fills this gap. We explored two factors that may affect NNSs’ contribution to the expressional and/or ideational aspects of joint content production with an NS: the order of turn-taking and the use of AI-powered editing tools. To unpack the effect of turn-taking order, we conducted an online experiment with 32 NS-NNS groups. Half of these co-writer groups followed a turn-taking order where an NNS acted first, while NSs acted ahead of NNSs in the other half. Our data revealed that NNSs had a low likelihood to edit the expressional aspect of the joint document regardless of the order of turn-taking. However, they were more inclined to edit the ideational aspect of the document when their editing turn occurred prior to an NS’s turn, as opposed to after. This contrast was accompanied with corresponding differences in participants’ self-reported coordination experience. Further, we found that all NNSs frequently leveraged AI-powered translators and paraphrasers to generate English expressions. This practice ended up causing unintended coordination issues that negatively impacted the interpersonal dynamics between co-writers. Due to a lack of awareness of NNSs’ interaction with tools, NSs had little clue to disentangle NNSs’ proficiency from the tools’ performances in producing fluent English. They also ran the risk of misinterpreting the tools’ outputs as full representations of an NNS’s voice and agency. Building upon these findings, we outlined implications for the design of future collaborative writing systems. In particular, we advocated for systems that can properly assess and promote each party’s contributions across the ideational, expressional, and lexical aspects. 

\nocite{*}

\begin{acks}
This work is supported by National Science Foundation, under grant \#1947929. We thank Qinxin Shu for her assistance, and Yongle Zhang, Jian Zheng, and Victoria Chang for their feedback. We also thank the ACs and anonymous reviewers for their valuable comments on earlier versions of this paper. 

\end{acks}


\bibliographystyle{ACM-Reference-Format}


\appendix
\section*{APPENDIX A. The Reader’s Letter Used for Each Writing Topic in the Task}

\textit{Topic A. Social Media}. “I use social media all the time. At first, seeing what my friends and families were up to made me quite happy even though they might be far away from me. However, there has been some changes recently. It just looks to me that everyone has a better life than I do, and I am nothing compared to them. But I should be happy for my friends, right? Where does my negative feeling come from? What can I do to overcome it? I hope I can receive some advice.” \\[3pt]
\textit{Topic B. Remote Learning}. “I recently heard from several parents in my daughter’s cohort that their kids are taking online programs after school. There are online tutors assisting the kids with their homework or offering advanced classes. It makes me worried that my kid would fall behind because she has not been taking those programs. Am I overconcerned? Will the online programs put some students in unfair dis-advantages? I hope to hear guidance about what to do.” \\[3pt]
\textit{Topic C. Digital Privacy}. “My information is being collected online all the time, be it my employment records, computer browning history, or where I spent my money. While it sometimes makes my life easier, there is this issue of privacy. I am worried that I have little control over who can access my data and what they will do with it. Am I worrying too much? What is your opinion about life in this datafied society? I hope to have a better understanding of this.”\\

\newpage
\section*{APPENDIX B. Examples of Rhetorical Pieces Receiving Different Types of Edits }

\centering
\begin{small}
    \begin{tabular}{p{5.3cm} p{5.3cm} p{1.9cm} }
\toprule
    \raggedright Content of a given rhetorical piece by the end of a turn & Content of the same rhetorical piece by the end of the previous turn & Context of this piece \\
    \midrule
    No edits & & \\
    “Libraries provide children with safe spaces to do homework, access the internet, and many of them offer extracurricular activities and programming. In recent years many libraries have expanded access to take-home devices as well as mobile internet hotspots for people with library cards.”   &  “Libraries provide children with safe spaces to do homework, access the internet, and many of them offer extracurricular activities and programming. In recent years many libraries have expanded access to take-home devices as well as mobile internet hotspots for people with library cards.”  & The co-writer group argued for the possibility of learning using public facilities.  \\
    & & \\
    Expressional edits only & & \\
    “\textbf{A study} published in 2016 found that social media can mirror our interactions with friends in real life: if we have a positive exchange, it will boost our self-esteem, and if we have a negative exchange it could lead to anxiety and depression. \textbf{It’s as if our brains experience physical pain sometimes on social media. You scroll Twitter, Facebook, or Instagram, feel envious, and simultaneously your brain feels a sort of physical pain.}”  & “\textbf{One study} published in 2016 found that social media can mirror our interactions with friends in real life: if we have a positive exchange, it will boost our self-esteem, and if we have a negative exchange it could lead to anxiety and depression. \textbf{someone say, your brain had taken envious as physical pain. If you scroll the twitter, facebook, or instagram, you feel envious and your brain get physical pain.}”  & The co-writer group argued for the potential harm of using social media to compare oneself with others.  \\
    & & \\
    Ideational edits only & & \\
    “Once you create a Nintendo account and enter your credit card number,  you can purchase games everywhere and anytime. In the past, I had to buy a prepaid card and enter the code number. I thought it was so tiresome. \textbf{Moreover, the same account can be shared by different game consoles, and gameplay videos can be shared on SNS.}”   &  “Once you create a Nintendo account and enter your credit card number, you can purchase games everywhere and anytime. In the past, I had to buy a prepaid card and enter the code number. I thought it was so tiresome.”  & The co-writer group argued for the benefit of storing personal information online.  \\
    & & \\
    Expressional and ideational edits & & \\
    “\textbf{Paper might actually behoove young learners in lieu of a screen. That is, although the latest digital technology may seem advantageous for students, it doesn’t necessarily mean that the information processed through this medium improves reading ability.}”  & “\textbf{In fact, some studies have shown that paper books are easier to understand and remember.}”  & The co-writer group argued for the possibility of learning without going digital.  \\
    \bottomrule
    & & \\
\end{tabular}
\end{small}
\raggedright

\newpage
\section*{APPENDIX C. NNS-11's Use of AI-Powered Tools at Their Editing Turn}

\centering

\begin{small}
    \begin{tabular}{p{3cm} p{3cm} p{3cm} p{3cm} }
    \toprule
    Japanese input written in each trial & English meaning of the Japanese input & English text outputs generated by the translation tool & Final sentence used in the article \\
    \midrule
    \begin{CJK}{UTF8}{min} 自分自身が好きになった \end{CJK}& \textit{I have come to like myself.} & \textit{I fell in love with myself.} [Google Translate] & \textit{I am confidence of myself.} \\
     & & & \\
     \begin{CJK}{UTF8}{min} 自分が好きになった \end{CJK} & \textit{I have come to like me.} & \textit{I fell in love with myself.} [Google Translate] & \\
     & & & \\
     & & \textit{I came to love myself.} [Yahoo Chiebukuro] & \\
     & & & \\
    \begin{CJK}{UTF8}{min} 自分自身が好きになった \end{CJK} & \textit{I have come to like myself.} & \textit{I fell in love with myself.} [Google Translate] & \\
     & & & \\
     & & \textit{Always remember, the most important thing is to love yourself first.} [Yahoo Chiebukuro] & \\
     & & & \\
     \begin{CJK}{UTF8}{min} 自分に自信がついた \end{CJK} & \textit{I have gained confidence in myself.} & \textit{I gained confidence in myself.} [Google Translate] & \\
     & & & \\
     \begin{CJK}{UTF8}{min} 自信を持つ \end{CJK} & \textit{Have confidence.} & \textit{Have confidence.} [Yahoo Chiebukuro] & \\
    \bottomrule
    & & \\
    
\end{tabular}

\end{small}

\raggedright

\end{document}